\documentclass[12pt]{iopart}
\newcommand{\be}{\begin{equation}}
\newcommand{\ee}{\end{equation}}
\newcommand{\bea}{\begin{eqnarray}}
\newcommand{\eea}{\end{eqnarray}}

\begin{document}

\title{Spin dynamics of the $S=1/2$ antiferromagnetic zig-zag ladder
with anisotropy}

\author{P. D. Sacramento}

\address{Departamento de F\'{\i}sica and  CFIF, Instituto Superior
T\'ecnico, Av. Rovisco Pais, 1049-001 Lisboa, Portugal}

\author{V. R. Vieira}

\address{Departamento de F\'{\i}sica and  CFIF, Instituto Superior
T\'ecnico, Av. Rovisco Pais, 1049-001 Lisboa, Portugal}

\date{\today}
%\maketitle

\begin{abstract}
We use exact diagonalization and the modified Lanczos method
to study the finite energy and finite momentum spectral weight 
of the longitudinal and transverse spin excitations of the anisotropic
zig-zag ladder. We find that the spin excitations form continua of
gapless or gapped spinons in the different regions of the phase
diagram. The results obtained are consistent with a picture previously
proposed that in the anisotropic case there is a transition from a gapped
regime to a gapless regime, for small interchain coupling. In this
regime we find a sharp low-energy peak in the structure function for the
transverse spin excitations, consistent with
a finite stiffness.
\end{abstract}

%Uncomment for PACS numbers title message
%\pacs{00.00, 20.00, 42.10}

\maketitle

\section{Introduction}
Recently \cite{Coldea2d} it has been suggested that a two dimensional spin
system, $Cs_2CuCl_4$, has an excitation spectrum that can
be described, similarly to the one dimensional
case \cite{Heisenberg}, by a continuum originated from pairs of spin
$1/2$ spinons. 
The standard one dimensional Heisenberg
model is known to have fractional states where the usual spin $1$
magnons are replaced by pairs of deconfined spin $1/2$ topological excitations
called spinons \cite{Muller}. The characteristic low excitation
energy coherent peaks that appear in the dynamical susceptibility are in
this case replaced by a continuum.
This property has been verified experimentally for several quasi-one
dimensional spin $1/2$ systems like CPC \cite{CPC}, $KCuF_3$ 
\cite{Nagler,Tennant} and copper benzoate \cite{Dender} where a 
description in terms of a nearest-neighbor Heisenberg model 
is assumed to apply.

Real materials are however neither strictly one-dimensional nor the interactions
are of the simple Heisenberg nearest-neighbor form. Originally $Cs_2CuCl_4$
was taken as a quasi-one-dimensional system \cite{Coldea} but a more careful
estimate of the inter-chain parameters revealed that they are of the same order
as the intrachain interactions \cite{Coldea2d}. 
The interlayer coupling is estimated to be two orders of magnitude smaller
implying that the system is essentially two-dimensional. It forms a triangular
lattice in the copper ($Cu^{2+}$, $S=1/2$) planes and constitutes therefore
a frustrated system.

Frustrated systems have attracted considerable interest. Using a large-$N$
bosonic expansion it has been predicted \cite{Sachdev1} that the presence of frustration
may counteract the staggered fields responsible for confinement \cite{Watanabe}
and lead to deconfinement of the spinons. In a non-frustrated system it is
known that the low energy modes are spin-$1$ magnons. The presence of frustration
leads to a non-collinear order parameter \cite{Sachdevb} which can be parametrized
by three real scalar numbers. The representation of the spin operators
in terms of bosonic operators contains an hidden internal gauge symmetry under which
the scalar link fields transform either as charge $+2$ or charge $+1$ scalars \cite{Sachdev1}. 
It has been predicted a long time ago that if the charge $+2$ scalars condense into a
Higgs phase \cite{Fradkin} then the unit charges are not confined and the spinons
remain free \cite{Sachdevb}. However the Higgs phase is just one of the possible
phases predicted to occur in the frustrated two-dimensional lattice. The $Cs_2CuCl_4$
system provides a first experimental example of a two-dimensional system with
a spectrum consistent with the existence of spinons.

One may think of a simpler system like the zig-zag ladder, where only two chains 
are coupled, and study the excitation spectrum of this system as a first 
step towards understanding the two-dimensional triangular lattice obtained in the
limit when several ladders are coupled to each other. In usual spin ladders
an infinitesimal coupling between chains leads to a behavior qualitatively
different from the $1d$ case. Therefore it is interesting to see if in the case of
the zig-zag ladder there is a qualitative difference from the single-chain case.
It has been found before that if the next-nearest-neighbor (nnn) interaction is small
enough the system still behaves qualitatively as the single-chain case. Therefore
it seems reasonable that for small couplings we might find similar features
characteristic of the single chain, in particular a spinon originated spectrum.
The main question to be answered is what happens for large couplings where the role
of the nnn coupling is important. We will show that indeed for all couplings
a spinon description holds. A direct comparison with the experimental results for
$Cs_2CuCl_4$ would require a full $2d$ calculation but our results for the zig-zag
ladder give a first indication that the nnn coupling does not lead to a coherent
energy spectrum.

In the context of a frustrated two-leg ladder similar to the zig-zag ladder
\cite{Allen} it has been shown that the spinons survive as the elementary
excitations in a spontaneously dimerized ground state but become massive.
A local $Z_2$ symmetry related to independent translations by one lattice
spacing on each chain is spontaneously broken and leads to a non-vanishing
dimerization for strong enough frustration. This symmetry breaking leads
to kinks as elementary excitations which are massive. These kinks
have been shown to have $S=1/2$ and therefore at least two must be created.

In general if a term removes explicitly the $Z_2$ degeneracy betweeen the two
dimerized ground states this leads to soliton confinement. An example is to
introduce explicit dimerization in the Hamiltonian.
The role of explicit dimerization has also been addressed in the context of spin-Peierls
systems like $CuGeO_3$ \cite{Hase}, $NaV_2O_3$ \cite{Isobe} or
$Cu(NO_3)_2 2.5 D_2 O$ \cite{Xu}. The excitation spectrum of these systems
is however considerably different from the chains without dimerization.
The spectrum is gapped and the lowest energy excitations are coherent spin-$1$
magnon peaks \cite{Bouzerar} that separate from the continuum that appears
at higher energies. The spectrum of these systems is actually closer to integer
spin chains \cite{Haldane} or to spin-ladders \cite{Dagotto}. The effects
of explicit dimerization have received renewed attention recently \cite{Dimer}.

Another source of interest in the zig-zag ladder is that
it has been proposed that in the anisotropic case 
incommensurate quasi-long-range spin correlations should be observed. Also 
a gapless chiral phase has been predicted to occur \cite{Nersesyan}.
In this work we focus our attention on the combined effects of a next-nearest
neighbor frustrating interaction and of anisotropy. 
In a two-dimensional non-frustrating system the spectrum is coherent and composed
of magnons. The addition of frustrating terms may lead to deconfined spinons.
In the zig-zag ladder 
when the nnn frustrating coupling is absent
the system has a spectrum determined by gapless spinons (it is the case of
the simple Heisenberg chain). Adding frustration it is expected that the
spinons will remain deconfined. In the isotropic case where the effect of the
nnn coupling is to dimerize the spinons are massive at sufficiently strong nnn coupling. 
In the anisotropic case at strong enough nnn couplings we expect the system to have a 
transition from the intermediate dimer phase to a gapless phase where we expect the 
deconfined spinons to be massless. 

\newpage
\section{Hamiltonian}
 
The anisotropic zigzag ladder is defined by the Hamiltonian
\bea
H & = & \frac{1}{2} J_{1}^{XY} \sum_{i} \left( S_{i}^{+} S_{i+1}^{-} +
S_{i}^{-} S_{i+1}^{+} \right) + J_1^{z} \sum_{i} S_{i}^{z} S_{i+1}^{z} \nonumber \\
& + & \frac{1}{2} J_{2}^{XY} \sum_{i} \left( S_{i}^{+} S_{i+2}^{-} +
S_{i}^{-} S_{i+2}^{+} \right) + J_2^{z} \sum_{i} S_{i}^{z} S_{i+2}^{z}.
\eea
The spin operators refer to spin $S = 1/2$ states, while
the summation $i=1,...,N$ runs along the ``rib'' of the zig-zag ladder.
We shall  parameterize the interactions by the 
coupling parameter $j=J_2^{XY}/J_1^{XY}$ and
by the anisotropy parameter
$J_1^z/J_1^{XY}=\Delta=J_2^z/J_2^{XY}$.
(The isotropic case reduces to $j=J_2/J_1$ and $\Delta=1$.) 
We will set $J_1^{XY}=1$ henceforth. 
The nearest-neighbor Heisenberg
chain with anisotropy corresponds to both the weak-coupling
($J_1 = 0$) and to the strong-coupling ($J_2 = 0$) limits of the
zig-zag ladder. The spectrum is gapless for
the case of $XY$ anisotropy, $|\Delta| \leq 1$,
as shown by the Bethe ansatz \cite{Baxter}.
The excitation spectrum
consists of spin-$1/2$ particles dubbed spinons. 
Since flipping one spin 
represents a spin-$1$ excitation,
%yields a change of $\Delta S_z=1$ 
the spinons can only be created in pairs. 
Therefore the conventional
spin $1$ magnons are deconfined into 
spin-$1/2$ spinons that propagate incoherently.

The isotropic case has been studied 
before \cite{Okamoto,Eggert,Nathalie,Chitra,White96,Allen2}
as a function of the coupling parameter, $j = J_2/J_1$.
As $j$ increases, the system goes from gapless (single chain)
to a dimer phase and then to a spiral phase, where the structure factor
has a maximum at a momentum $\pi/2<q<\pi$. 
The system has a spin gap 
in these last  two phases,
and it  therefore only displays short-range order. 
In the limit that
the intra-chain interaction is much larger than the inter-chain interaction
($j \rightarrow \infty$) the two chains decouple and a gapless single chain
behavior is recovered. 
It has been argued that this  only happens, strictly speaking,
at $j= \infty$: the spin gap becomes exponentially 
small as $j$ grows, but it remains non-vanishing \cite{White96}.
Recently, on the other hand,
 it has been proposed that
incommensurate quasi-long-range spin correlations should be observed
if easy-plane ($XY$)
anisotropy is included in the zigzag ladder \cite{Nersesyan}. 
This is argued
to be due to the presence of a ``twist'' term
that results from the inter-chain
interaction. It has been proposed that there is one gapless mode and one mode
with a  gap in the regime of strong $XY$ anisotropy in the 
inter-chain coupling. Another prediction
of this work is the existence of
spontaneous local 
spin currents.  This,  however,
has been refuted in ref. \cite{Kaburagi}.
Also, other recent numerical  work~\cite{Aligia}
has failed to confirm the gapless nature of the groundstate
in the anisotropic $XY$ case at weak interchain coupling. 
Recent Density Matrix Renormalization Group (DMRG) 
results~\cite{Hikihara} suggest, however, that the zig-zag ladder 
does indeed show a gapless chiral
phase as predicted in ref. \cite{Nersesyan}.

%%%%%%%%%%%%%%%%%%%%%%%%%%%%%%%%%
Also, recently an analysis of the exact properties of
such finite systems was carried out, looking at various correlation functions and
the structure of the spectrum both in the isotropic and the
anisotropic cases \cite{Vieira}. 
The spin stiffness of the zig-zag ladder was calculated, 
and it was found evidence 
for a gapless regime at weak coupling  that survives
the  thermodynamic limit
in the case of	$XY$ anisotropy. 
This was also concluded looking at the level crossings to detect the
phase transition between the two regimes \cite{Vieira} using a previously proposed
procedure to detect the dimer transition at strong interchain coupling
\cite{Okamoto}. The same method was also recently used in reference \cite{Somma}.

In this work we will study the structure function and the spectral 
weight of the spin excitations
both for the longitudinal and the transverse correlations. 
Our results are consistent
with previous conclusions that there is a transition to a gapless regime
at weak coupling if anisotropy is present. The results indicate a
continuum of gapless excitations in the transverse correlations in the $XY$ case,
and a continuum of gapped excitations in the isotropic case.

\section{Spectral weight}

The structure function is defined by the overlap of two states coupled either
by the longitudinal or the transverse spin operator \cite{Muller},
\be
S_{\mu \nu}(q,\omega) = \frac{1}{N} \sum_{l,R} e^{i q R} 
\int_{-\infty}^{\infty} dt e^{i \omega t} <S_l^{\mu}(t) S_{l+R}^{\nu}(0)>
\ee
where $\mu$ and $\nu$ are cartesian components. At zero temperature we obtain therefore
\be
S_{\mu \mu}(q,\omega) = \sum_{\lambda} M_{\lambda}^{\mu} \delta (\omega +
E_G -E_{\lambda} ) \delta (q+k_0 -k_{\lambda} )
\ee
where $E_G$ is the groundstate energy, $E_{\lambda}$ is the energy of an excited
state, $\omega$ is the excitation energy (energy difference to the groundstate)
and $q$ is its momentum (momentum difference to the momentum of the groundstate
$k_0$) and the spectral weight is defined by
\be
M_{\lambda}^{\mu} = 2 \pi \left| <G | S^{\mu}(q) |\lambda > \right|^2.
\ee
where $S^{\mu}(q)$ is the Fourier transform of the spin operator. We will 
calculate the structure functions $S_{+-}(q,\omega)$ and $S_{zz}(q,\omega)$
which probe the transverse and the longitudinal spin excitations, respectively.

The single chain case was studied before both by Bethe ansatz \cite{Heisenberg}
and using numerical diagonalization of small systems \cite{Muller}. The continuum
of excitations is contained in the thermodynamic limit between two
lines: the bottom one is the single-spinon dispersion, $\epsilon_l(q)$, and
the upper one is the maximum energy resulting from the combined effect of
two spinons, $\epsilon_u(q)$. In the isotropic case ($\Delta=1$)
the lines are defined by 
\bea
\frac{\epsilon_l(q)}{J_1} & = & \frac{\pi}{2} \left| \sin (q) \right| \nonumber \\ 
\frac{\epsilon_u(q)}{J_1} & = & \pi \left| \sin \left( \frac{q}{2} \right) \right| 
\eea
and in the $XY$ case ($\Delta=0$) are defined by
\bea
\frac{\epsilon_l(q)}{J_1^{XY}} & = & \left| \sin (q) \right| \nonumber \\ 
\frac{\epsilon_u(q)}{J_1^{XY}} & = & 2 \left| \sin \left( \frac{q}{2} \right) \right| 
\eea

In the $XY$ case the longitudinal structure function can be calculated exactly 
\cite{Niemeijer} since this system is equivalent to free spinless 
fermions and it is given by \cite{Muller}
\be
S_{zz}(q,\omega) = 2 \frac{\Theta(\omega-\sin (q)) \Theta(2 \sin (q/2)-\omega )}{
\sqrt{4 \sin^2(q/2)-\omega^2}}
\ee
In the isotropic case there is no exact solution but M\"{u}ller et al. \cite{Muller}
proposed an ansatz that fits very well both numerical results for small systems
and various experimental results where a description in terms of a single chain
is expected to hold. The M\"{u}ller ansatz is
\be
S_{zz}(q,\omega) = \frac{A}{\sqrt{\omega^2 - \epsilon_l^2(q)}} \Theta(\omega
-\epsilon_l(q)) \Theta(\epsilon_u(q)-\omega)
\ee
where $\Theta$ is a step function and $A$ a constant \cite{Luther,Muller}. 
This function diverges
at the lower boundary while in the $XY$ case it diverges at the upper
boundary \cite{Careful}. At momentum $\pi$ the divergence is stronger and it diverges
as $S_{zz} \sim \omega^{-1}$.

In the thermodynamic limit the structure function eq. (3) can be written
as a product \cite{Muller}
\be
S_{\mu \mu}(q,\omega) = M^{\mu}(q,\omega) D(q,\omega)
\ee
where $M^{\mu}(q,\omega)$ is the continuum limit of the spectral weight originating
in the overlaps eq. (4) and $D(q,\omega)$ is the density of states. In the isotropic
and in the $XY$ case the density of states is finite and nearly constant close to 
the low-energy threshold and it diverges at the upper threshold. On the other 
hand $M^{\mu}(q,\omega)$ is constant in the $XY$-case and it diverges at the 
lower threshold in the Heisenberg case. The structure function as a consequence 
diverges in the lower threshold for the Heisenberg chain and it diverges 
in the upper threshold in the $XY$-case \cite{Muller}. For any finite system 
the density of states is a set of delta functions at the excitation energies. 

Using field theory it is also possible to determine the dependence
of the transverse structure function close to the lower threshold. 
In the single chain case the transverse function is given by \cite{New}
\be
S_{+-}(q, \omega) \sim \frac{1}{\left( \omega^2 - \epsilon_l^2(q)
\right)^{3/4} }; j=0, \Delta=0
\ee
and therefore
\be
S_{+-}(\pi,\omega) \sim \omega^{-3/2}
\ee
However the finite energy structure function is not known analytically.

The ladder case is more involved. We will use 
exact diagonalization of finite systems
together with the modified Lanczos method \cite{Gagliano}.

\section{Numerical results}

Let us begin by recalling the quantum numbers of the groundstate
as a function of the size $N$
for the $S = 1/2$ zig-zag antiferromagnet.
Periodic boundary conditions are imposed throughout.
The groundstate is 
a spin singlet in general due to the antiferromagnetic
interactions.
The system has three well defined regimes: ({\it a}) strong-coupling,
({\it b}) intermediate coupling and ({\it c}) weak-coupling.
Consider the isotropic case first.
For strong enough coupling between chains, $j = J_2/J_1 < 1/2$, 
it
has either momentum $\pi$ for $N=4n+2$
or  momentum $0$ for $N=4n$. 
For intermediate couplings ($j>1/2$), on the other hand,
the momentum oscillates between $0$
and $\pi$ as a function of the coupling 
parameter $j$ and of the system size $N$ \cite{Tonegawa87}. 
There are several points along $j$ 
in this regime
where the corresponding  energy levels
for these two momentum values cross. 
The groundstate is degenerate at these points,
and this is reflected by  peaks in the 
dimer correlation function \cite{Vieira,Majumdar}. 
Such level crossings grow in number as the system size grows,
and this indicates that the two singlet states in question
are in fact degenerate in the
thermodynamic limit.  
By the Lieb-Schultz-Mattis theorem,~\cite{Lieb} 
this is consistent with a 
spin gap in the excitation spectrum that survives the
thermodynamic limit  in the weak-coupling regime $j > 1/2$. 

The spectrum of
the anisotropic $S = 1/2$ $XXZ$ zig-zag ladder has also been studied
previously 
in the strong-coupling regime
up to the Majumdar-Ghosh line ($0 < j < 1/2$).~\cite{Nomura}
A gapless regime occurs for $XY$ anisotropy  $\Delta \le 1$ and
strong coupling
$j<j_{c1}(\Delta)$;  an Ising antiferromagnet along the rib
of the zig-zag that shows a spin gap in the excitation spectrum
occurs for $\Delta>1$ and $j<j_{c1}(\Delta)$,
and a dimer phase regime that also has a spin gap
exists at  $j>j_{c1}(\Delta)$ and any $\Delta$.
The line $j = j_{c1}(\Delta)$
separates the gapless phase from the dimer phase for $\Delta \le 1$,
while it
separates the dimer phase from the (Ising) N\'{e}el phase for $\Delta>1$.
The line at $\Delta=1$ and $j<j_{c1}$ separates the 
$XY$ gapless phase from the Ising phase.

It was found \cite{Vieira} that there is
a transition from the gapped intermediate coupling regime to a weak-coupling
gapless regime. In the intermediate coupling regime the two lowest
states are two states with $S_z=0$, of momenta $k=0,\pi$. There is a line,
$j_{c2}(\Delta)$, where the first excited state becomes a $S_z=\pm 1$,
$k=\pi/2$ state signalling the doubling of the periodicity and leading
to a gapless regime as confirmed from the spin stiffness tensor highest
eigenvalue \cite{Vieira}.
The curve $j_{c2}(\Delta)$ shown in Fig. 4 of ref. \cite{Vieira} separates a 
spin-gap (dimer) phase from a gapless phase
at small interchain couplings. As expected, the value of $j_{c2}$ grows   
near the isotropic point.
(It should tend to $j=\infty$ 
at $\Delta = 1$ according to White and Affleck, \cite{White96} but
finite-size effects gave a finite value).

Let us now analise the spectral weight eq. (4) at various points
in the phase diagram parametrized by $j$ and $\Delta$. We will
focus our attention on two classes of parameters. We will consider
the isotropic case ($\Delta=1$) and the $XY$-case ($\Delta=0$) varying
in both cases the interchain coupling, $j$.

Let us begin with the single chain case for both values of $\Delta$ and let us
consider the particular case $N=16$. In Fig. 1 we show the lowest energy levels
(taking the groundstate as the zero of energy) for $S_z=0$ and $S_z=1$ and a
given momentum for the Heisenberg chain and the $XY$ chain. In the
Heisenberg case the states are organized into spin multiplets due to the
$SU(2)$ spin invariance. The groundstate is a spin singlet with momentum
zero. The first excited state is a spin triplet with momentum $k=\pi$ and
the next state is another spin singlet but with momentum $k=\pi$. In the
$\Delta=0$ case the first excited state is now a state with $S_z=1$ and momentum
$k=\pi$ and the next state is fourfold-degenerate with $S_z=0$ and momentum
$k=4 \pi /N$ or momentum $k=\pi$, or $S_z=2$ and the same momentum values
\cite{Vieira}. 

The structure function $S_{zz}$ only couples the groundstate
to states with $S_z=0$ and $S_T=1$. On the other hand $S_{+-}$ only couples
the groundstate to states with $S_z=1$ and $S_T=1$ ($S_T$ is the total spin).
In the isotropic
case $S_{+-}$ couples to a subset of the states probed by $S_{zz}$ while in
the anisotropic case the two functions probe different sets of states.

In Fig. 2a we show the spin excitations that contribute to $S_{zz}(q,\omega)$
for $N=16$ and for $\Delta=1$. As mentioned above the spectral weight of the spin
excitations decreases as we move away from the lower threshold. The states contained
in the region defined by $\epsilon_l(q)$ and $\epsilon_u(q)$ have a considerable
weight while those at higher energies have a much smaller weight \cite{Muller}.
The spectral weight of these higher states will vanish in the thermodynamic limit.
Also other states contained in the continuum have very small weights. The continuum
is therefore well defined by the set of states with highest spectral weight.
In Fig. 2b we show the structure function for the Heisenberg chain. The
delta functions at the excitation energies have been given a finite
width both in frequency and momentum for better visualization.

In Fig. 3 we show the states with non-vanishing spectral weight for the longitudinal
and the transverse structure functions for the $XY$ chain. The spectral weight is
uniform inside the continuum defined by eqs. (6).
The spectrum of the transverse excitations is however 
different. The lower spinon dispersion is well described by the single-spinon
dispersion; particularly close to $q=\pi$ the gap is already rather small
for such a small system. At higher energies the spectral weight is 
considerably spread.

As we introduce the next-nearest-neighbor interaction the spectrum remains gapless
for all $\Delta$ if $j$ is small. In the intermediate coupling regime ($j \sim 1$)
the system becomes gapped. In Fig. 4 we show the a) spectral weight and b) 
the structure function for $j=1$ in the isotropic case. The states with $S_z=0$
and momenta $k=0,\pi$ are nearly degenerate \cite{Vieira}. The next excited 
state is a spin triplet
with momentum $k=\pi /2$ which in the thermodynamic limit will have a gap to the
groundstate. Accordingly the spectral weight shows a gap with a continuum above it
indicative of massive spinons. In the anisotropic case the two lowest 
states are the same as in the isotropic case but the next excited state is a 
$S_z=1$, $k=\pi /2$ state \cite{Vieira}.
The next state is a $S_z=0$, $k=\pi$ singlet. The low value of
the gap signals the near level crossing that for $N=16$ 
occurs around $j=1.2$ leading
to a gapless regime \cite{Vieira}. In Fig. 5 we show
the spectral weights and the structure functions
for the longitudinal and the transverse spin excitations. In the case
of $S_{zz}$ the lowest gap is at $k=\pi$, while for $S_{+-}$ the lowest gap
is at $k=\pi /2$. It is also clear that the spectrum is quite sharp at $k=\pi /2$
in the transverse spin function.

As we increase $j$ further the sharpness of the gapless transverse 
mode at $k=\pi /2$
becomes stronger. In Fig. 6 we show the structure function 
for the isotropic case and in Fig. 7 the same function
for the anisotropic case at $j=2$. 
The Goldstone
mode predicted to occur in the anisotropic case for the transverse spin excitations
clearly singles out.

We also consider the finite-size dependence of the low-energy excitations
for the longitudinal and the transverse spectral functions as a function of the 
system size using results
from exact diagonalizations and the modified Lanczos method. We consider system
sizes up to $N=24$. 
The results extrapolate to the single-spinon
dispersion curve for the various values of $\Delta$ and $j$. In particular
we consider the anisotropic case.

In Fig. 8 we show the lowest energy states for a) $S_z=0$ and b) $S_z=1$
as a function of momentum for the $XY$  case for $j=2$.
The spectrum clearly shows the doubling
of the lattice cell with a significant low energy mode at $q=\pi/2$,
particularly in the transverse correlations ($S_z=1$) where once again as
the system size increases the gapless nature of the spectrum is evident as
$j$ grows (weak interchain couplings).

In the single-chain case the longitudinal spectrum can be obtained
considering the two-spinon curves (assuming non-interacting spinons) via the usual
procedure
$E(q)=\epsilon(k_1)+\epsilon(k_2)$
where $q=k_1+k_2$, $E(q)$ is the two-spinon curve and $\epsilon(k)$ is the
single-spinon dispersion curve. These two limiting curves define
the region of the continuum spectrum.
The transverse excitations in the anisotropic case probe however
a different set of states as can be seen for example from the single-chain
$XY$ case shown in Fig. 3c. The high energy part of the spectrum shows
that the interactions between the spinons can not be ignored (remember
that in the Jordan-Wigner transformation from a $XY$ chain to spinless
fermions, the transverse correlations involve the presence of strings).
Therefore the two-spinon rule requires a proper treatment of the spinon
interactions. The zig-zag ladder case is still more involved 
particularly for the transverse excitations. In any case the continuum 
is clearly visible.

%\newpage
\section{Conclusions}

The finite-energy and finite-momentum structure function provides a direct
way of analysing the excitation spectrum. Previously the structure factor
was analysed in the isotropic case \cite{Chitra,White96,Vieira}. The structure
factor is obtained integrating the structure function over frequency at
a fixed momentum. In the isotropic case the peak in the structure factor
shifts from $q=\pi$ to $q=\pi/2$ when the spiral phase is reached as the
Majumdar-Ghosh point is crossed. In the anisotropic case on the other hand
it has been predicted that in the limit of very weak interchain coupling
an incommensurate gapless chiral phase should be observed \cite{Nersesyan}.
However, for finite systems it is difficult to detect the incommensurability
since the shift from commensurability is very small \cite{Aligia}.

In this work we have analysed the structure function itself for the zig-zag
ladder as a function of anisotropy ($\Delta$) and the interchain coupling
($j$). The results show that in general the excitations are gapless or
gapped spinons that have to be created in pairs as for the single-chain
case. In the $XY$ case ($\Delta=0$) as $j$ grows it is clear that a gapless 
mode in the transverse excitations arises in agreement with previous results from field
theory \cite{Nersesyan} and with previous numerical results \cite{Vieira}
obtained analysing the stiffness and the level crossings. 
In the isotropic case ($\Delta=1$) the spectrum is a continuum of massive spinons.
These results may be relevant to understand the ladder limit in the context of the
recent experimental results on the
two-dimensional system $Cs_2CuCl_4$ \cite{Coldea2d}. 

After this work was completed we became aware of a preprint
\cite{Okunishi} where using M\"{o}bius boundary conditions it is shown
that in the isotropic case in the strong and intermediate coupling 
cases the spectrum may be described by a continuum resulting
from two-spinon scattering, in agreement with the general conclusions
of our paper.

%\newpage
\ack
We acknowledge several discussions with Jos\'{e} Rodriguez and
with Alexander Nersesyan.
This research was partially supported by the Program PRAXIS
XXI under grant number 2/2.1/FIS/302/94.

%%%%%%%%%%%%%%%%%%%%%%%%%%%%%%%%%%%%%%%%%%%%%%%%%%%%%%%%%%%%%%%%%%%%%%%%%%%%

\newpage
\bf
\begin{center}
Figure Captions
\end{center}
\rm

\vspace{\baselineskip}
\noindent
Fig. 1- Excitation energies from the exact diagonalization of a $N=16$
chain as a function of momentum for $S_z=0$ and $S_z=1$ for the Heisenberg
chain and the $XY$ chain.

\vspace{\baselineskip}
\noindent
Fig. 2- a) Excitation energies of the states that contribute to the longitudinal
spectral weight for the Heisenberg chain. The solid lines are the exact Bethe
ansatz results for the thermodynamic limit. The numerical results are obtained
via exact diagonalization of a $N=16$ system. The color code of the points is
the following: the points correspond to states with a spectral
weight that is i) $M(q,\omega)>1$ (black), ii) $1>M(q,\omega)>0.1$ (red) 
and iii) $0.1>M(q,\omega)>0.01$ (green). The same color codes are 
used in the remaining figures. In Fig. 2b we show the structure function.
The momentum is shown in units of $\pi$.
The vertical scale is in arbitrary units. 

\vspace{\baselineskip}
\noindent
Fig. 3- Excitation energies of the states that contribute to the a) longitudinal
and c) transverse spectral weight of the XY chain. The solid lines in Fig. 3a are
the exact Bethe ansatz results. We also show the structure function for the
b) longitudinal and d) transverse excitations.

\vspace{\baselineskip}
\noindent
Fig. 4- a) Excitation energies obtained from the exact diagonalization of a $N=16$
ladder in the isotropic case and $j=1$ that contribute to the spectral
weight and b) the structure function. 

\vspace{\baselineskip}
\noindent
Fig. 5- a) Excitation energies of the 
states that contribute to the a) longitudinal
and c) transverse spectral weight in the anisotropic case and $j=1$.  
We also show the structure function for the b) longitudinal and d) 
transverse excitations.

\vspace{\baselineskip}
\noindent
Fig. 6- Structure function of a $N=16$ ladder in the isotropic case and $j=2$. 

\vspace{\baselineskip}
\noindent
Fig. 7- Structure function for the a) longitudinal and b)
transverse excitations in the anisotropic case and $j=2$.  

\vspace{\baselineskip}
\noindent
Fig. 8- Lowest energy branch for $\Delta=0$ and $j=2$ for a) $S_z=0$ and
b) $S_z=1$ for different system sizes.


\begin{thebibliography}{100}
\bibitem{Luther} See also A. Luther and I. Peschel, Phys. Rev. B
{\bf 12}, 3908 (1975) where the exponent of the structure function for the Heisenberg
chain was obtained using the Luttinger model analogy.
\bibitem{Careful} Note however that in the case of the $XY$ chain the
high energy behavior is not well described by the Luttinger model
\cite{Luther}.
\end{thebibliography}
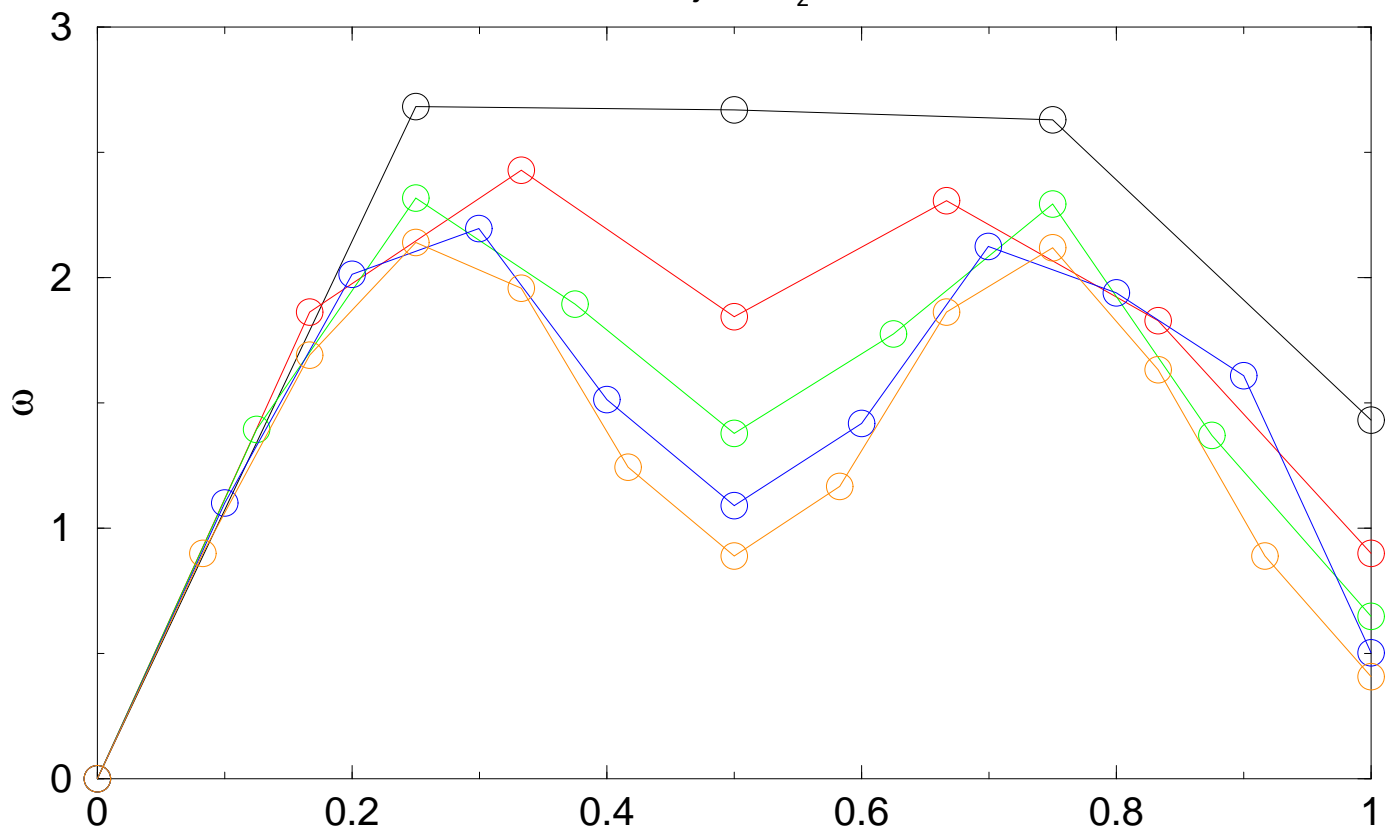
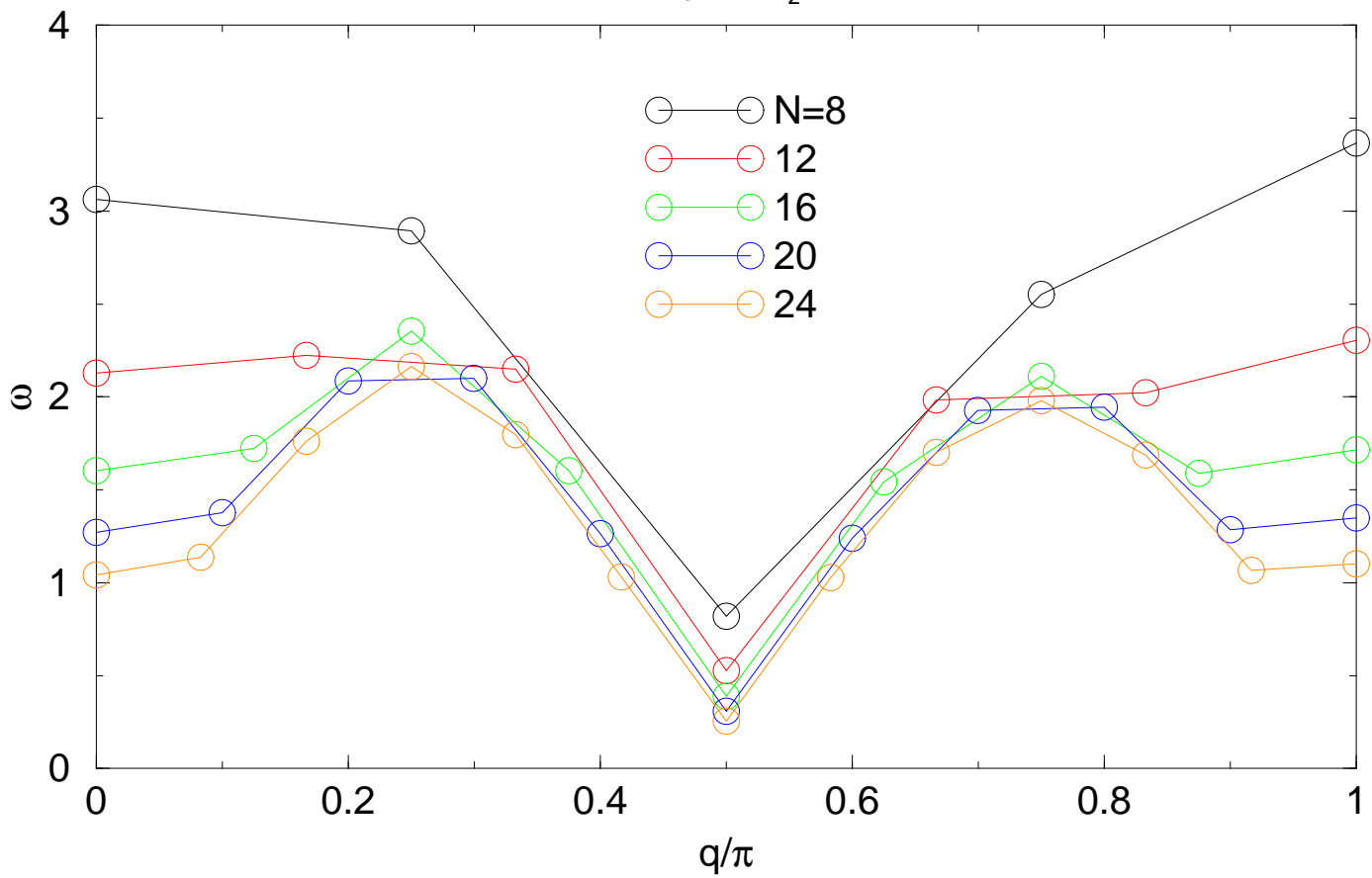
\end{document}